\documentclass[twocolumn,floatfix,superscriptaddress]{revtex4-1}

\usepackage{dsfont} % Typesetting the mathematical symbols for the natural numbers (N), whole numbers (Z), rational numbers (Q), real numbers (R) and complex numbers (C).
\usepackage{bm} % \bm to makfe bold fonts in math mode

\usepackage{bbold} %For identity
\usepackage{xcolor}
\usepackage{times}
\usepackage{physics}
\usepackage{graphicx,braket}
\usepackage{hyperref}
\hypersetup{colorlinks=true,linkcolor=blue,citecolor=blue}
\usepackage{amsmath,amssymb,amsfonts}
\usepackage{wrapfig} 

\usepackage{siunitx}

\usepackage{cleveref}% load last
\crefname{equation}{Eqs.}{Eqs.}
\Crefname{equation}{Equation}{Equations}% For beginning \Cref
\crefrangelabelformat{equation}{(#3#1#4--#5#2#6)}
\crefmultiformat{equation}{Eqs. (#2#1#3}{, #2#1#3)}{#2#1#3}{#2#1#3}
\Crefmultiformat{equation}{Equations (#2#1#3}{, #2#1#3)}{#2#1#3}{#2#1#3}
\usepackage{ amssymb }

\AtBeginDocument{%
    \newwrite\bibnotes
    \def\bibnotesext{Notes.bib}
    \immediate\openout\bibnotes=\jobname\bibnotesext
    \immediate\write\bibnotes{@CONTROL{REVTEX41Control,eprint=""}}
    \immediate\write\bibnotes{@CONTROL{%
    apsrev41Control,author="08",editor="1",pages="1",title="0",year="1"}}
     \if@filesw
     \immediate\write\@auxout{\string\citation{apsrev41Control}}%
    \fi
}%
\begin{document}

\flushbottom
\title{Squeezed lasing}

\author{Carlos S\'anchez Mu\~noz }
\affiliation{Departamento de Física Teórica de la Materia Condensada and Condensed
Matter Physics Center (IFIMAC), Universidad Autónoma de Madrid, Madrid,
Spain}
\affiliation{Clarendon Laboratory, University of Oxford, Parks Road, Oxford OX1 3PU, United Kingdom}
\email[Corresponding author: ]{carlossmwolff@gmail.com}
\author{Dieter Jaksch}
\affiliation{Clarendon Laboratory, University of Oxford, Parks Road, Oxford OX1 3PU, United Kingdom}

\newcommand{\down}{\op{g}{e}}
\newcommand{\up}{\op{e}{g}}
\newcommand{\downd}{\op{+}{-}} % Down for the Dressed basis
\newcommand{\upd}{\op{+}{-}}
\newcommand{\app}{a^\dagger}
\newcommand{\ssp}{\sigma^\dagger}
\newcommand*{\Resize}[2]{\resizebox{#1}{!}{$#2$}}%

\begin{abstract}
We introduce the concept of a squeezed laser, in which a squeezed cavity mode develops a macroscopic photonic occupation due to stimulated emission.  Above the lasing threshold, the emitted light retains both the spectral purity inherent of a laser and the photon correlations characteristic of a photonic mode with squeezed quadratures. Our proposal, which can be implemented in optical setups, relies on the parametric driving of the cavity and dissipative stabilization by a broadband squeezed vacuum. The squeezed laser can find applications that go beyond those of standard lasers thanks to the squeezed character, such as the direct application in Michelson interferometry beyond the standard quantum limit, or its use in atomic metrology.

\end{abstract}
\date{\today} \maketitle

\section{Introduction}

Squeezed states of light are one of most important resources in current quantum-optical technologies. Their main feature,  quadrature fluctuations below the shot-noise limit, has multiple applications in quantum metrology, with remarkable examples such as the enhanced sensitivities reported in gravitational-wave interferometers~\cite{aasi13a}. 
A particularly interesting possibility is the  application of squeezed states in atomic quantum metrology, where they have been proposed as a way to generate spin squeezing~\cite{Hald1999,Kuzmich1997,Hammerer2010,Appel2008,
Honda2008,Tanimura2006,Hetet2007} and enhance the sensitivity of atomic interferometers ~\cite{Agarwal1996,Szigeti2014,Pezze2018} and magnetometers~\cite{Wolfgramm2010,Wolfgramm2013,Horrom2012}. Efficient coupling between squeezed light and atomic transitions  requires the generation of narrowband squeezed states~\cite{Kim2018}. For this reason, one might argue that, in order to couple squeezed light with atoms, it would be ideal to have a source of squeezed light with the narrow linewidth characteristic of the lasers usually employed in atom optics. In lasers, such a narrow linewidth is a consequence of stimulated emission, but this mechanism is challenging to harvest for squeezing generation. The reason for this is that squeezed states are typically generated in an optical parametric oscillator (OPO) operating below the critical point. There, stimulated emission, which becomes dominant as one approaches criticality, is in fact a detrimental factor, because it turns the squeezed vacuum into a mixture of coherent states. Therefore, since squeezed states are produced \emph{below} the critical point of a driven-dissipative phase transition, contrary to lasers, they cannot benefit from the extremely narrow linewidth and diverging coherence times characteristic of states above the critical points. Formally, these desirable traits of a laser are the consequence of a closed Liouvillian gap~\cite{Kessler2012,Fernandez-Lorenzo2017,Minganti2018}.

In this work, we propose the implementation of a squeezed laser, introducing a mechanism of stimulated emission that generates coherent squeezed states retaining both the linewidth and coherence time characteristic of a laser, and the photon correlation properties of squeezed states.  This novel source of light can open new regimes of exploration of atomic physics with quantum states of light. We prove that the direct generation of bright squeezed states into a singe mode can also allow for new interferometry methods based on squeezed states without the need of mixing coherent states and squeezed vacua, thus preventing problems related to imperfect mode matching~\cite{Oelker2014}.

\begin{figure*}[t!]
\begin{center}
\includegraphics[width=0.8\textwidth]{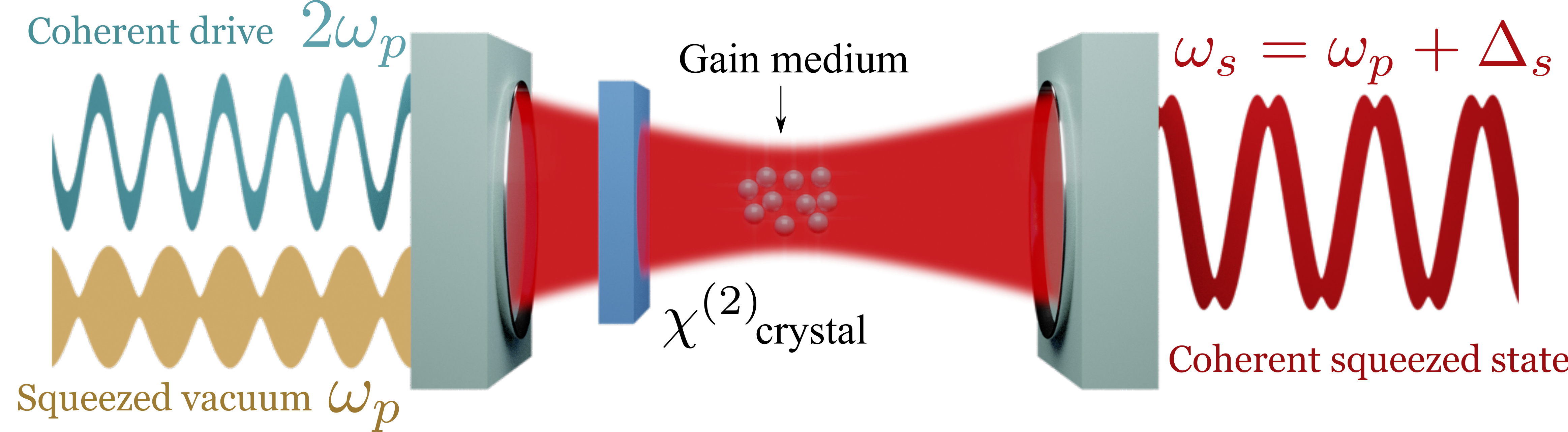}
\end{center}
\caption{Sketch of the proposed setup: a single cavity mode of frequency $\omega_c=\omega_p+\Delta_c$ (with $\Delta_c \ll \omega_p$) is parametrically driven through the down-conversion of pump photons of frequency $2\omega_p$ by a non-linear crystal $\chi^{(2)}$. The cavity includes a gain medium (e.g. an ensemble of two-level atoms) and is driven by a broadband squeezed vacuum centred at the frequency $\omega_p$ to stabilize lasing action. If the laser is imposed with a well-defined phase, the  output emission corresponds to a coherent squeezed state of frequency $\omega_s = \omega_p + \Delta_s$.}
\label{fig:setup}
\end{figure*}

\section{Model}
The mechanism that we propose requires driving the cavity that contains the gain medium by \emph{i)} a detuned parametric drive, which sets squeezed states as the natural photonic eigenstates of the system, and \emph{ii)} a resonant, broadband squeezed vacuum, which stabilizes the fluctuations  towards a squeezed vacuum.

Our model consists of a single cavity mode of frequency $\omega_c$ with bosonic annihilation operator $\hat a$, interacting with  $N$ two-level atoms with ground and excited energy levels $\{|g\rangle_i,|e\rangle_i\}$,  lowering operators $\hat\sigma_i \equiv |g\rangle_i\langle e|_i$, and transition frequency $\omega_\sigma$. The cavity is parametrically driven by a detuned drive of amplitude $\Omega_p$, achieved through the down-conversion of a coherent drive of frequency $2\omega_p$ into photon pairs at frequency $\omega_p$ (slightly detuned from the cavity frequency $\omega_c$) by means of a non-linear $\chi^{(2)}$ crystal inside the cavity  (see Fig.~\ref{fig:setup}). A resonant version of this type of parametric drive is the typical mechanism to generate a squeezed vacuum just below the OPO threshold. Here, the coupling to the atoms will provide a gain mechanism, amplifying the vacuum of this squeezed mode into a coherent squeezed state by stimulated emission, and yielding laser-like coherence times. Contrary to the case of stimulated emission in the OPO phase transition, this macroscopic population buildup occurs into a squeezed mode and preserves the squeezing properties rather than degrading them. As we prove later, the pump intensities required to turn normal lasing action into lasing into a squeezed mode are well within reach in current experimental platforms.

In a frame rotating at a frequency $\omega_p$, the total Hamiltonian reads ($\hbar = 1$)
\begin{equation}
\hat H = \Delta_c \hat a^\dagger \hat a + \frac{\Omega_p}{2}(e^{-i\theta}{{\hat a}^2 }+\mathrm{H.c}) + \sum_{i=1}^N  \Delta_\sigma \hat \sigma_i^\dagger \hat \sigma_i+g(\hat a^\dagger \hat\sigma_i + \hat a \hat\sigma_i^\dagger) ,
\end{equation}
with $\Delta_i \equiv  \omega_i-\omega_p$, and $\theta$ the phase of the coherent drive. We have assumed a Jaynes-Cummings type of light-matter coupling, requiring the rotating-wave approximation (RWA) $g\ll \omega_c,\omega_\sigma$.  
The purely photonic part of the Hamiltonian can be diagonalized by a Bogoliouvov transformation corresponding to a unitary squeezing operator $\hat S(re^{-i\theta})=\exp[r(e^{i\theta} \hat a^2 - e^{-i\theta} {\hat{a}^{\dagger 2}})/2]$, with $r\equiv\ln[(1+\alpha)/(1-\alpha)]/4$, and $\alpha \equiv \Omega_p/\Delta_c$, so that $\hat a\rightarrow \hat a_s\cosh r-\hat a_s^\dagger e^{-i\theta}\sinh r$, where $\hat a_s$ denotes the annihilation operator in the new, squeezed basis. The Hamiltonian then approximately becomes
\begin{equation}
\hat H \approx \Delta_s \hat a^\dagger_s \hat a_s + \sum_{i=1}^N  \Delta_\sigma \hat \sigma_i^\dagger \hat\sigma_i+\tilde g(\hat a_s^\dagger \hat \sigma_i + \hat a_s \hat \sigma_i^\dagger),
\label{eq:Hamiltonian-squeezed}
\end{equation}
where $\Delta_s \equiv \Delta_c\sqrt{1-\alpha^2}$ and $\tilde g \equiv g\cosh r$. In the last step we have performed a new RWA under the requirement that the collective coupling remains small compared to the effective free frequencies, $\sqrt{N} g \sinh r \ll \Delta_s, \Delta_\sigma$. Although this requirement can in principle always be satisfied for any $r$ by increasing both $\Delta_c$ and $\Omega_p$ so that the ratio $\alpha$ remains constant, it will ultimately be limited by the realistic impositions on the driving amplitude, $\Omega_p$, and by the condition that $\Delta_c$ remains smaller than the free spectral range of the cavity, $\omega_\mathrm{FSR}$. 
The resulting light-matter coupling rate $\tilde g$ is exponentially enhanced with respect to the bare coupling by the factor $\cosh r$: this enhancement and its implications in different setups have been proposed and discussed in several works in previous years~\cite{Lemonde2016,Zeytinoglu2017,Leroux2018,Qin2018}. Here, we explore another consequence of this type of coupling: the possibility of developing a macroscopic photonic phase in the squeezed cavity mode $a_s$ through a lasing mechanism.

To consider the possibility of squeezed lasing, we must work with a driven-dissipative model in which the previous Hamiltonian is supplemented by Lindblad operators that describe incoherent driving of the atoms and dissipative decay of cavity photons and atomic excitations. As it has been discussed in previous works~\cite{Lemonde2016,Qin2018}, a cavity coupled to a thermal reservoir behaves as a cavity coupled to a squeezed reservoir in the squeezed basis. This effective squeezed noise is detrimental for lasing action in the squeezed cavity mode, and needs to be removed. One can achieve this by driving the system with a broadband squeezed vacuum with an opposite squeezing angle and a properly tuned squeezing parameter~\cite{Qin2018}, which can be obtained from the output of an OPO of frequency $\omega_p$ and with a linewidth much larger than the cavity decay rate~\cite{gardiner_book00a}. 

Let us define $\kappa$ as the cavity photon loss rate through the mirror that couples to the squeezed photonic reservoir (i.e., the output of the driving OPO). We will also consider other sources of photon loss  with a total associated decay rate $\eta \kappa$, where $\eta$ is a dimensionless factor. We consider these extra losses to be smaller or similar to $\kappa$, so that $\eta \lesssim 1$. These other sources of loss can, for instance, encompass intra-cavity losses, or losses through the second mirror in a two-sided cavity, as shown in Fig.~\ref{fig:setup}. The total decay rate in the cavity, considering these extra sources of loss, is $\kappa(1+\eta)$.

As we elaborate in the Supplemental Material, considering driving by a squeezed vacuum with squeezing parameter $r_e e^{i\theta_e}$, a lasing master equation can be obtained by setting
\begin{equation}
r_e = r + \frac{1}{2}\text{asinh}[\eta \sinh(2r)] \approx r + \frac{1}{2}\eta\sinh(2r)
\end{equation}
(the last equality being valid for $\eta \ll \sinh 2r$) and  $\theta_e =\pi-\theta$. This yields the following master equation for the atoms-cavity system
\begin{multline}
\partial_t{\hat\rho} = -i[\hat H,\hat \rho] + \frac{\kappa}{2}(1+\eta+N_s)D_{\hat a_s}[\hat \rho]
+\frac{\kappa}{2}N_s D_{\hat a^\dagger_s}[\hat \rho]\\ +\sum_{i=1}^N \left[\frac{P}{2}D_{\hat \sigma_i^\dagger}[\hat \rho] + \frac{\gamma}{2}D_{\hat \sigma_i}[\hat \rho]\right],
\label{eq:master-equation}
\end{multline}
where $H$ is given by Eq.~\eqref{eq:Hamiltonian-squeezed}, $P$ and $\gamma$ are, respectively, the incoherent pumping and spontaneous emission rates of the atoms, and $N_s$ is an effective thermal photon number given by
\begin{equation}
N_s = \frac{1}{2}\left[2 \eta \sinh^2 r + \sqrt{1+\eta^2 \sinh^2(2r)} -1 \right].
\end{equation}

\begin{figure}[t!]
\begin{center}
\includegraphics[width=0.99\columnwidth]{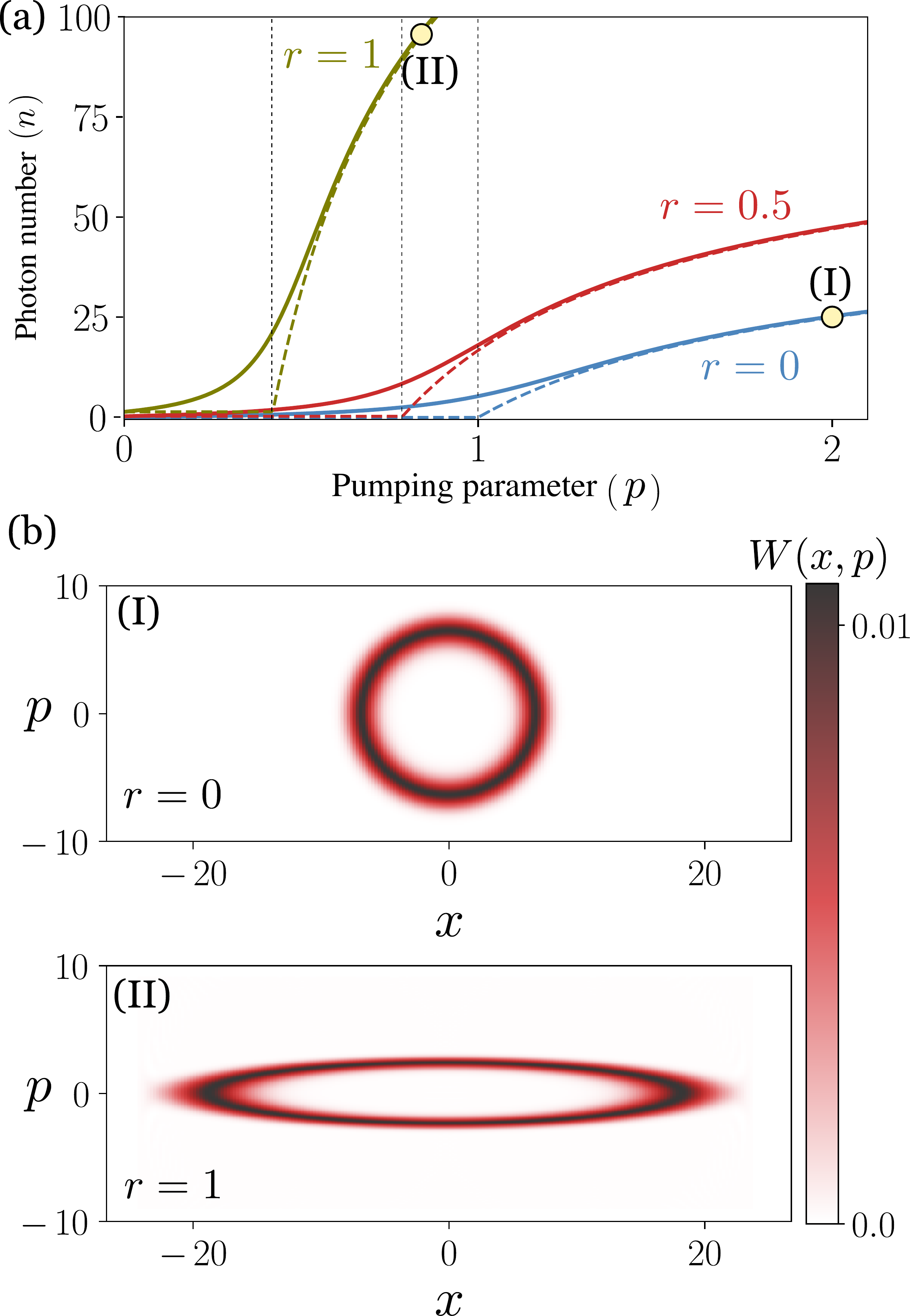}
\end{center}
\caption{(a) Lasing phase transition in terms of photon number versus pumping parameter, for different values of the squeezing parameter $r$. Saturation photon number set as $n_q=50$. Vertical lines mark the critical points, which are reduced for increasing $r$. Solid lines: exact numerical result. Dashed lines: mean field prediction. (b) Wigner function of the steady state corresponding to points (I) and (II) in panel (a), with $r=0$ and $1$ respectively, and $\theta=\pi$.  }
\label{fig:phase-transition}
\end{figure}
%ç
The master equation \eqref{eq:master-equation} describes a extensively studied model of a laser in contact with a thermal bath~\cite{gardiner_book00a}, and therefore will yield lasing action in the squeezed cavity mode. The finite effective temperature of the bath, which tends to drive the system towards a thermal state of mean photon number $N_s$, has in principle a detrimental effect for lasing. Nevertheless, \textcolor{black}{as we show in the Supplemental Material}, provided that the coupling to the squeezed reservoir remains the principal mechanism of photon loss, i.e. $\eta \lesssim 1$, the thermal photon number can be kept $N_s \sim 1$ or lower, and thus its detrimental effect for the generation of a macroscopic, coherent population of photons is negligible. Therefore, we will set for simplicity $N_s=0$ for the rest of the work.
  
Let us define the atom decay $\tilde\gamma\equiv(P+\gamma)/2$, the squeezed cooperativity $ C_s=2 g_s^2/(\tilde\gamma \kappa)$, the pump parameter $p_s=NC_s (P-\gamma)/(P+\gamma)$ and the saturation parameter $n_0 = \tilde\gamma^2/2 g_s^2$. A mean-field description of the Heisenberg equations of motion given by the ansatz $\langle \hat a_s^\dagger \hat \sigma\rangle = \langle \hat a_s^\dagger\rangle \langle \hat \sigma\rangle$ predicts a photon population $n_s \equiv  \langle \hat a_s^\dagger \hat a_s\rangle$ featuring a dissipative phase transition at $ p_s=1$,
\begin{equation}
n_s= n_0(p_s-1) H(p_s-1) 	,
\end{equation}
with  $H(x)$  the Heaviside step function.  
Let us now consider this expression in terms of both the ``bare'' cooperativity $C=2g^2/\kappa \tilde\gamma$ and pumping parameter $p=p_s C/C_s$. Given that $C_s = C \cosh^2 r$,  the threshold condition $p_s=1$ can be written as $p=1/\cosh^2 r$. Therefore, one of the consequences of the parametric driving of the system is a reduction of the lasing pumping threshold by a factor $1/\cosh^2 r$, as can be see in Fig.~\ref{fig:phase-transition}(a). 

Although many practical situations will require to consider a finite atomic decay $\gamma\neq 0$, for the sake of simplicity we will take the limit of a long-lived atomic transition $\gamma\rightarrow 0$, since the essential physics that we will discuss remains unchanged. By doing so the parameters above reduce to $p_s=N C_s$ and $n_0 = n_q/ N C_s$, where $n_q=NP/2\kappa$ is the photon saturation number. The mean-field result thus predicts a phase transition at $p_s=1$, with a photon population
\begin{equation}
n_s= n_q \left( 1-\frac{1}{p_s}\right) H(p_s-1).
\end{equation}
The photonic population in the bare cavity mode $n \equiv  \langle \hat a^\dagger \hat a \rangle $ can be expressed in terms of $n_s$ by undoing the squeezing transformation, which gives:
\begin{equation}
n = n_s \cosh(2r)+\sinh^2 r.
\label{eq:na_squeezed}
\end{equation}
These results can be taken to the limit of the one-atom laser $N=1$, where $p_s = C_s$. While this limit is of practical and fundamental interest~\cite{McKeever2003,Boozer2004,Florescu2004,Astafiev2007,Dubin2010,Navarrete-Benlloch2014}--- particularly due to the thresholdless behaviour around the phase transition---we will adopt it just for numerical convenience, since it will allow us to provide numerically exact results for the macroscopic cavity field. We will not focus on any particular feature of the one-atom limit, since we shall be concerned with the deep-lasing regime, where the results are expected to be similar to the $N\gg 1$ case, converging to the mean-field prediction. There will be therefore no loss of generality.

\section{Stationary state}
Figure~\ref{fig:phase-transition}(a) depicts exact result for the stationary photon number $n$ of the one-atom squeezed laser versus the pumping parameter $p$, here equal to the cooperativity $C$, for various values of $r$,  together with the prediction of the mean-field ansatz. The exact solution depicts the well-known thresholdless behaviour of the one-atom laser~\cite{McKeever2003}, and it converges to the mean-field prediction at large values of $p$. Despite the fact that the gain consists of a single atom, the system still features a notion of thermodynamic limit, which is reached when the photon saturation number, set by the ratio between $P$ and $\kappa$, tends to infinity, $n_q \rightarrow \infty$. In this limit, the photon number shows the non-analytical behaviour around the critical point predicted by the mean-field calculation.

Figure~\ref{fig:phase-transition}(b) shows the Wigner function of the lasing states corresponding to $r=0$ and $r=1$, at the points I and II marked in Fig.~\ref{fig:phase-transition}(a). These stationary states display a characteristic, non-gaussian annular shape. Well within the lasing phase, a lasing state is well approximated by a mixture of coherent states $|\sqrt{n_s} e^{i\varphi}\rangle$ with a fixed amplitude $\sqrt{n_s}$, mixed over all possible phases,
\begin{equation}
\hat \rho_s = \frac{1}{2\pi}\int_0^{2\pi}d\varphi \,  |\sqrt{n_s} e^{i\varphi}\rangle \langle\sqrt{n_s}e^{i\varphi}|.
\end{equation}
This mixed stationary state reflects the effect of phase diffusion in the limit of infinite time, and it is a common property of all quantum models of a laser. However, these models also predict, as we will show, that the timescale of phase diffusion is extremely slow and related to the inverse of the linewidth, yielding the characteristically long coherence times of the laser. In our case, notably, this lasing state is developed in the squeezed basis. To recover the stationary photonic state of cavity mode in the original basis, $\hat \rho_a$, we must apply a squeezing transformation, giving
\begin{equation}
\hat \rho_a = \frac{1}{2\pi}\int_0^{2\pi}d\varphi \,  S(r e^{i\theta})|\sqrt{n_s} e^{i\varphi}\rangle \langle\sqrt{n_s} e^{i\varphi}|S(r e^{i\theta})^\dagger.
\label{eq:rho_a}
\end{equation}
This mixture of displaced-squeezed states gives the squeezed annular shape in phase space displayed at the bottom panel of Fig.~\ref{fig:phase-transition}(b).  To the best of our knowledge, a similar squeezed-lasing state has only been considered in a previous proposal in the microwave domain Ref.~\cite{Navarrete-Benlloch2014}, which required a fast modulation of qubit energies in circuit quantum electrodynamics.  Our proposal allows to implement this novel type of lasing in the optical domain, or any other system where a broadband squeezed vacuum and efficient down-conversion mechanisms are available. 

\begin{figure}[t!]
\begin{center}
\includegraphics[width=0.99\columnwidth]{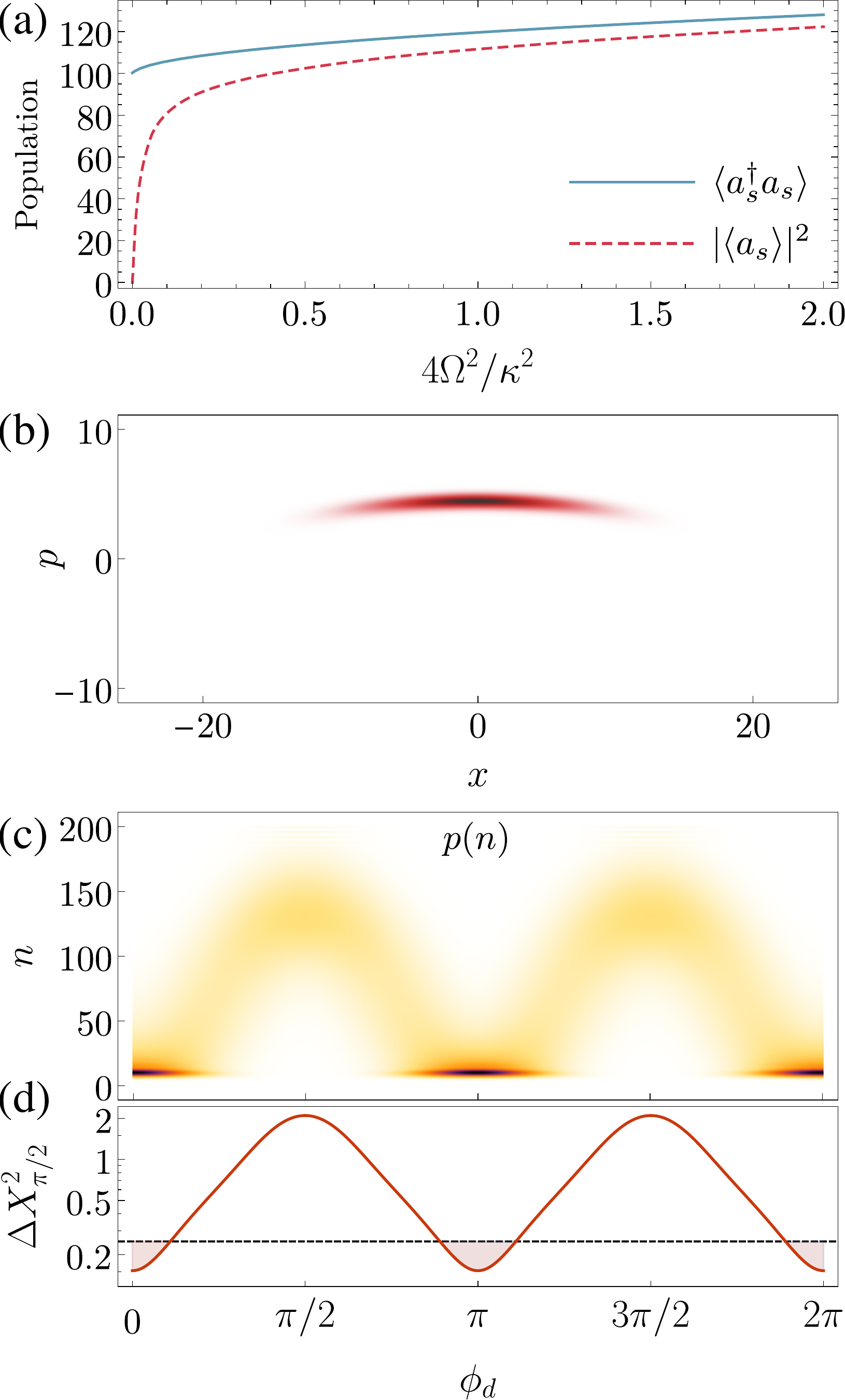}
\end{center}
\caption{Enforcing a well defined phase. (a) Stationary photon population in the squeezed basis, compared to the square of the mean field, versus the amplitude of the coherent drive. Parameters: $n_q = 200$, $C=2$ (b) Wigner function of the steady-state of the squeezed laser with the addition of a small drive to enforce a phase, with the same parameters as in (a) and $r=1$. (c) Photon number distribution and (d) quadrature fluctuations versus the phase of the driving field, with the same parameters as in (a) and $r=0.7$. Squeezing can be achieved with drives with phase $\phi_d = 0,\pi$.   }
\label{fig:symmetry-breaking}
\end{figure}

\subsection*{Symmetry broken solutions}
\label{sec:symmetry-breaking}
The phase-diffused stationary state does not reflect the multiple ways in which, just as in a conventional laser, the system may select a phase that would then remain stable for a extremely long  time.  For instance, it is expected that a specific phase will be spontaneously selected under an homodyne measurement~\cite{Bartolo2017}. Alternatively, as in any other physical system with a symmetry, one can add an external perturbation, such as a small external field, that breaks the symmetry~\cite{Fernandez-Lorenzo2017}. To elaborate further on this point, Fig.~\ref{fig:symmetry-breaking} summarizes the effect of adding a small coherent drive with a well-defined phase, described by the  Hamiltonian term $\hat H_\mathrm{drive} = \Omega(\hat a_s e^{-i\phi_d}+\hat a_s^\dagger e^{i\phi_d})$. The stationary population of photons that would be established in the squeezed basis solely by the action of this drive is $n_d =4\Omega^2/\kappa^2$.  As shown in Fig.~\ref{fig:symmetry-breaking}(a), a drive that would induce a small population of a just a few photons, $n_d \sim 1$, is enough to force a phase in a lasing state of $\sim 100$ photons in the squeezed basis. This is evidenced by the fact that $|\langle a_s \rangle|^2$ goes from zero (no well defined phase) to $\sim \langle \hat a^\dagger_s \hat a_s \rangle$, indicating that the annular distribution in phase space turns into a coherent state. In the standard cavity basis, the coherent state turns into a squeezed coherent state, as shown in Fig.~\ref{fig:symmetry-breaking}(b). We note that these states, which make up the mixture in Eq.~\eqref{eq:rho_a}, are of the form%
\begin{equation}
\hat\rho_\varphi = |\xi,\alpha\rangle\langle\xi,\alpha|
\label{eq:rho_varphi}
\end{equation}
with  $|\xi,\alpha\rangle\equiv\hat S(\xi)\hat D(\alpha)|0\rangle$ and $\xi = r e^{i\theta}$, $\alpha = \sqrt{n_s}e^{i\varphi}$. This is different from a coherent squeezed state, since $\hat S(\xi)\hat D(\alpha)\neq\hat D(\alpha)\hat S(\xi)$. This difference is important since, in the latter case, the photon number $n = \sinh r^2 + |\alpha|^2$ is completely independent of the relative phase $\theta-\varphi$, whereas in the former it is not. This can be easily understood noting that a squeezed coherent state is in fact a coherent squeezed state, but with a coherent component $\gamma$ different from $\alpha$ and dependent on $r$ and $\theta$, $\hat S(\xi)\hat D(\alpha)=\hat D(\gamma)\hat S(\xi)$. with $\gamma = \alpha\cosh r - e^{i\theta}\alpha^* \sinh r$. When $\theta-\varphi=\pi$, we find the maximum amplification of the photon number  $n = \sinh^2 r + |\alpha|^2 e^{2r}$ (the average over all possible relative phases recovers Eq~\eqref{eq:na_squeezed}). Symmetry-broken states imposed by an external drive display this oscillatory dependence of photon number on the relative phase between coherent component and squeezing angle, as depicted in Fig.~\ref{fig:symmetry-breaking}(c-d), showing how the degree of squeezing, photon statistics and mean photon number can be controlled through the drive phase. 

 In the absence of any external drive, all these symmetry-broken states can be can be considered to be metastable, and they are related to the fact that the Louvillian gap closes in the lasing regime, as we discuss below.

\section{Emission properties}
Having discussed the implementation of a squeezed laser, we now analyse its quantum-optical properties with the aim of establishing how the essential properties from both lasing and squeezing are unified in this light source. We  turn our attention to two fundamental characteristics that define both lasers and squeezed states: the spectral distribution of the emission, and the fluctuation properties. We will assume that there is an output channel of the laser different from the channel where the squeezed drive is being applied and analyze the properties of such output, e.g.. the emission from a second mirror, as sketched in Fig.~\ref{fig:setup}. This will allow us assume that the input of this channel is in a vacuum state, so that standard input-output relations apply.

\subsection{Spectrum of emission}
The spectrum of emission in the stationary limit $t\rightarrow\infty$ is  defined in terms of the Wiener-Khintchine formula~\cite{eberly76a}
\begin{equation}
S(\omega) = \lim_{t\rightarrow \infty}\frac{1}{\pi n} \Re\int_0^\infty d\tau\, e^{i\omega \tau} \langle \hat a^\dagger (t) \hat  a(t+\tau) \rangle.
\label{eq:spectrum}
\end{equation}
To relate the spectrum of the squeezed laser to that of a normal laser, it is convenient to rewrite the correlation function in terms of those of the squeezed mode. Writing $c\equiv \cosh r$, $s\equiv \sinh r $, we obtain:
%
%\begin{widetext}
\begin{multline}
\langle \hat a^\dagger (t) \hat a(t+\tau) \rangle = c^2\langle \hat a^\dagger_s(t)\hat a_s(t+\tau)\rangle+ s^2 \langle \hat a_s(t)\hat a_s^\dagger (t+\tau)\rangle  \\
+ cs \left[e^{i\theta}\langle \hat a_s^\dagger (t)\hat a^\dagger_s(t+\tau) \rangle + e^{-i\theta}\langle \hat a_s(t)\hat  a_s(t+\tau) \rangle  \right].
\label{eq:correlator-transformations}
\end{multline}
%\end{widetext}
%

\begin{figure}[b]
\centering
\includegraphics[width=0.99\columnwidth]{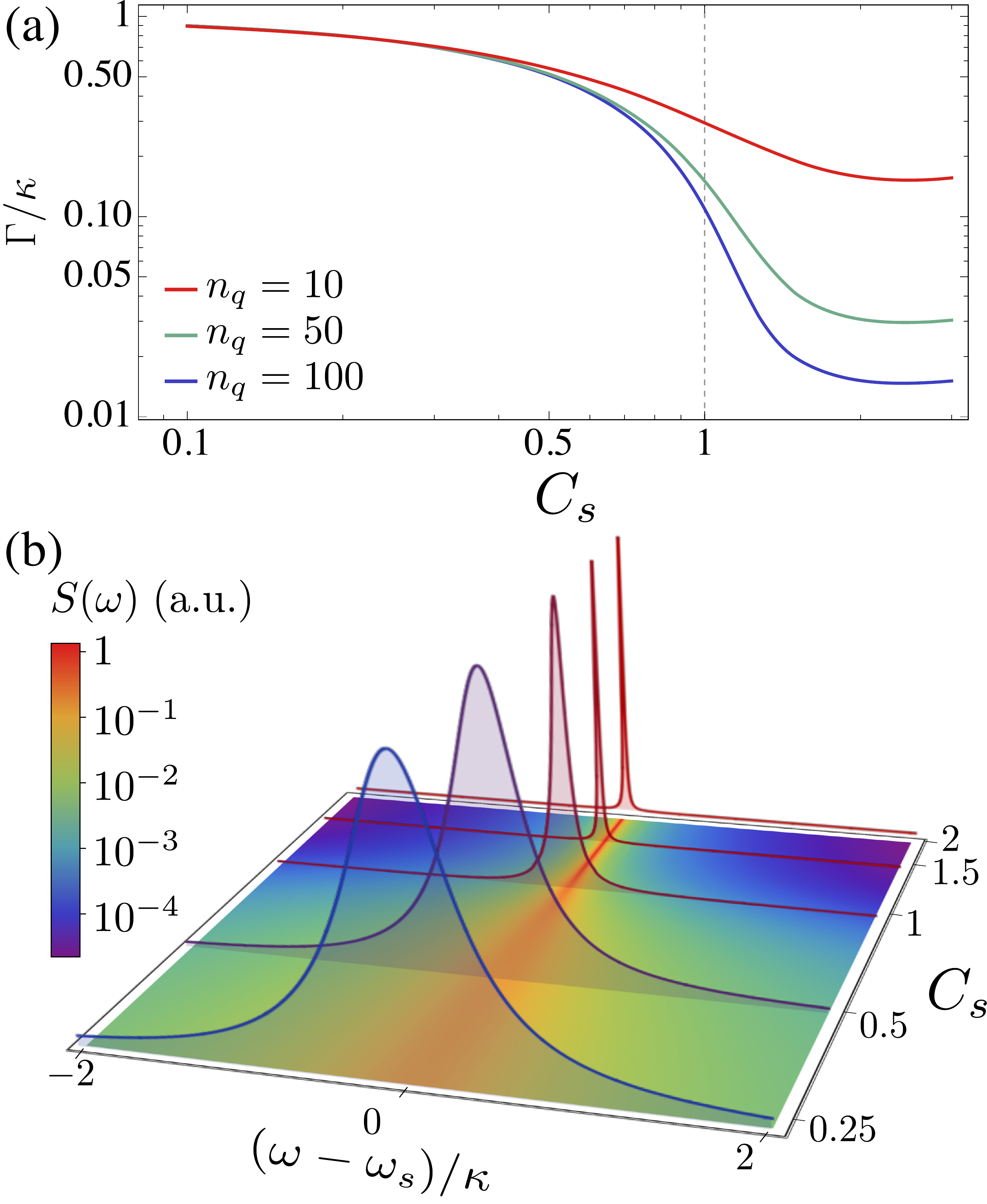}
\caption{Line narrowing of the emission spectrum in a squeezed laser. (a) Emission linewidth (related to phase diffusion rate) versus cooperativity $C_s=p_s$ for different values of saturation photon number $n_q$. The lasing phase transition occurs at $C_s = 1$.  This result depends on $r$ only through $C_s$. (b) Spectrum versus cooperativity $C_s$ for saturation photon number $n_q=100$.  Frequency defined with respect to the squeezed mode frequency $\omega_s$.}
\label{fig:spectrum}
\end{figure}  
From the master equation~\eqref{eq:master-equation}, we find that, in the stationary regime, any expectation value carrying phase information, such as $\langle \hat a_s ^2\rangle $, vanishes due to phase diffusion. This allows us to disregard the term proportional to $cs$ in Eq.~\eqref{eq:correlator-transformations}. Since the first two terms give similar contributions to the spectrum, we obtain $S(\omega) \propto S_s(\omega)$, where $S_s(\omega)$ is given by Eq.~\eqref{eq:spectrum} with $\hat a$ substituted by $\hat a_s$. Since the dynamics of the squeezed mode $\hat a_s$, given by Eq.~\eqref{eq:master-equation}, is that of a standard laser, $S_s(\omega)$---and subsequently $S(\omega)$---corresponds to the spectrum of a laser with photon number $n_s$ and cooperativity $C_s$. We can therefore conclude that the squeezed laser will inherit the spectral properties of a standard laser, including the extreme line narrowing developed in the lasing phase, with a linewidth that in the thermodynamic limit will be given by  $\Gamma = \kappa C_s/4 n_s$~\cite{gardiner_book00a}. The timescale of phase diffusion is given by $\Gamma^{-1}$. The spectrum will be sharply peaked around the frequency of the squeezed mode $\omega_s \equiv \omega_p + \Delta_s$. Well within the lasing regime, this linewidth is inversely proportional to the saturation photon number $n_q$, and therefore will vanish in the thermodynamic limit $n_q \rightarrow \infty$. This trend is confirmed by numerical calculations of the spectrum of the squeezed laser, see Fig.~\ref{fig:spectrum}, where we show the values of $\Gamma$ extracted from the spectrum versus the effective cooperativity of the squeezed mode, $C_s$ (with the lasing transition at $C_s=1$) and increasing values of the saturation photon number $n_q$. 

Writing the lasing master equation Eq.~\eqref{eq:master-equation} in terms of the Liouvillian superoperator, $\dot \rho = \mathcal L \rho$, the spectral linewidth $\Gamma$ can equivalently be estimated from the Liouvillian gap, e.g., the eigenvalue $\lambda_1$ of the Liouvillian superoperator $\mathcal L $ with the second largest real part, so that $\Gamma \approx - 2\Re\{\lambda_1\}$ (this relationship is discussed in more detail in~\cite{SanchezMunoz2019}).  Line-narrowing in lasing can therefore be understood as the result of a second-order driven-dissipative phase transition in which the Liouvillian gap is closed~\cite{Kessler2012,Minganti2018}. Being so intertwined with the spectral properties of the Liouvillian, it is not surprising that the linewidth of the spectrum of emission is unchanged by a unitary transformation. As we will see, this is not the case for observables that quantify the correlation amongst the emitted photons. 

\subsection{Quadrature squeezing}
Most technological applications of squeezed states aim to exploit the reduced fluctuations in one of the quadratures of the field for metrological purposes~\cite{Demkowicz-Dobrzanski2015,Lvovsky2015,Dowling2015,Schnabel2017}. It is therefore pertinent to ask whether the squeezed-lasing steady state given by Eq.~\eqref{eq:rho_a} shows any degree of quadrature squeezing. 
Defining the quadratures $\hat X_\phi =\hat  a e^{-i\phi} + \hat a^\dagger e^{i\phi}$,  the squeezed quadrature with minimal fluctuations is the one with angle $\phi = \theta/2$. In the stationary state, fluctuations of this quadrature are given by
\begin{equation}
\langle \Delta \hat X_{\theta/2}^2 \rangle = e^{-2r}(2n_s+1).
\label{eq:squeezing-stationary}
\end{equation}
\begin{figure}[b]
\centering
\includegraphics[width=0.99\columnwidth]{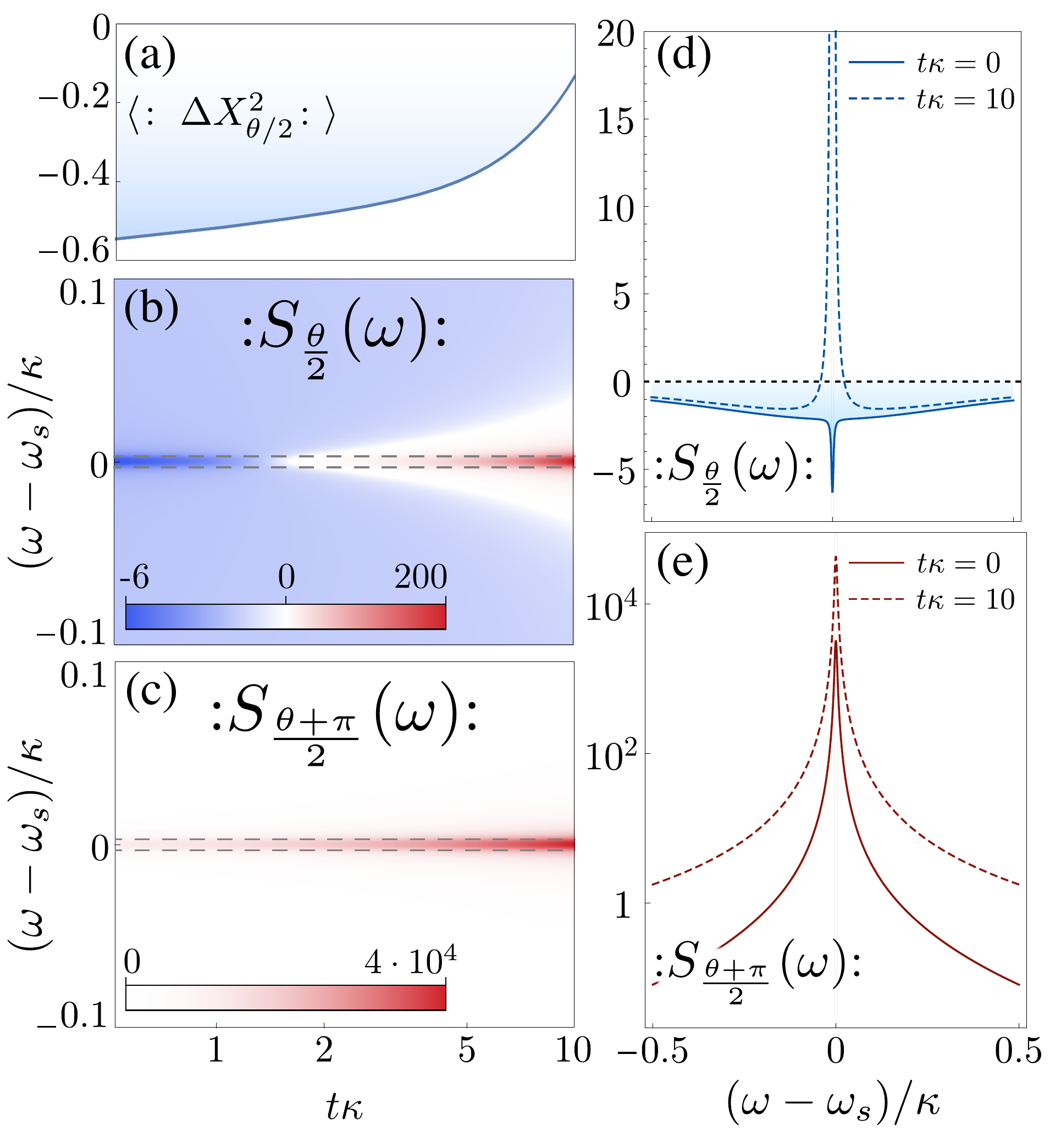}
\caption{Time evolution of squeezing.  (a) Normalized quadrature variance and (b-c) squeezing spectrum versus time  (negative values indicate fluctuations below the shot noise). (d-e) Represent cuts of (b-c) at initial and final time. We observe survival of squeezing for times much longer than cavity lifetime. Parameters, $n_q = 250$, $C_s=2$, $\varphi=(\theta+\pi)/2$, $r=0.45$.}
\label{fig:spectrum-squeezing}
\end{figure}  
As a consequence of the phase-diffusion, the exponential reduction of the quadrature fluctuations typical of a pure squeezed state is now multiplied by a factor $2n_s +1$, which will yield squeezing, i.e. $\langle \Delta \hat X_{\theta/2}^2 \rangle < 1$, whenever $n_s < (e^{2r}-1)/2$. This squeezing condition cannot hold in the deep-lasing regime $n_s \gg 1$. However, we emphasize that Eq.~\eqref{eq:squeezing-stationary} only pertains to stationary, phase-diffused solutions of the problem, which are reached on a timescale $\tau_c = 1/\Gamma $ that, as demonstrated above, tends to diverge in the lasing regime. 
Moreover, we have also shown that a very small external field suffices to lock a phase of the laser in the squeezed basis. 
Therefore, as it is usually done in the standard lasing scenario~\cite{Molmer1997}, it is meaningful to consider the properties of such  metastable symmetry-broken states with a well-defined phase $\varphi$, which we can describe as pure coherent squeezed states. Based on our discussion in Sec.~\ref{sec:symmetry-breaking}, we assume that this phase can be imposed into the system. Once this is done, the vanishing Liouvillian gap implies that the corresponding symmetry-broken state $\hat \rho_\varphi$ in Eq.~\eqref{eq:rho_varphi} will remain stable with a diverging lifetime, without the need of any external field. We will now further verify these claims and investigate the squeezing properties of such symmetry-broken states.

Let us consider the squeezing along a given quadrature $\hat X_\phi$ for the state $\hat \rho_\varphi$, which depends on both $\theta$ (set by the phase of coherent drive) and $\varphi$. The quadrature $\hat X_{\theta/2}$ remains the one with minimal fluctuations regardless the value of $\varphi$, and for an optimum phase $\varphi = (\theta\pm \pi)/2$, one recovers the degree of squeezing characteristic of a squeezed vacuum, 
\begin{equation}
\langle \Delta \hat X_{\theta/2}^2 \rangle = e^{-2r},
\end{equation}
meaning that the squeezed laser indeed operates as a source of coherent squeezed states. The most standard way to assess quadrature fluctuations in the light emitted by quantum-optical systems is its analysis in the spectral domain in terms of the power spectrum. In the particular case of homodyne measurements,  one generates difference photocurrents proportional to the expected value of a selected quadrature $\hat X_\phi^\mathrm{out}$ of the output field. The power spectrum of this signal contains contributions from quantum fluctuations that are encoded in the spectrum of squeezing~\cite{walls_book94a}, $S_\phi(\omega) = 1 + :\mathrel{S_\phi(\omega)}:$, with
\begin{equation}
:\mathrel{S_\phi(\omega)}: \equiv \int_{-\infty}^\infty \langle :\mathrel{\hat X_\phi^\mathrm{out}(t+\tau),\hat X_\phi^\mathrm{out}(t)}:\rangle e^{-i\omega \tau}\, dt,
 \end{equation}
where $\langle A,B\rangle = \langle AB\rangle - \langle A\rangle \langle B\rangle$ and $::$ denotes normal ordering. Negative (positive) values of this quantity indicate fluctuations below (above) the shot noise limit. While one usually considers the stationary limit $t\rightarrow \infty$, we will maintain the dependence on $t$ in order to be able to study the long-lived transient states that, as we have discussed, display the strongest squeezing features in our system.
Using standard input-output theory~\cite{walls_book94a}, we can write $:\mathrel{S_\phi(\omega)}:$ in terms of the intracavity operators as
\begin{multline}
:\mathrel{S_\phi(\omega)}:=2\kappa \int_{0}^{\infty}\cos(\omega \tau)\left[\langle \hat a^\dagger(t+\tau),\hat a(t)\rangle \right.\\
\left. + e^{-2i\phi}\langle \hat a(t+\tau),\hat a(t)\rangle + \mathrm{c.c.} \right].
\end{multline}
Figure~\ref{fig:spectrum-squeezing} summarizes the time evolution of the squeezing properties of an initial,  phase-locked state $\hat \rho_\varphi$, with $\varphi = (\theta+\pi)/2$. Our claim that these states are of a metastable nature are further supported by the time evolution of $\langle : \mathrel{\Delta \hat X^2_{\theta/2}}:\rangle$, see Fig.~\ref{fig:spectrum-squeezing}(a), which shows that the quadrature remains squeezed for a time much longer than the cavity lifetime $\kappa^{-1}$. As shown in Fig.~\ref{fig:spectrum-squeezing}(b-c), the profile of the spectrum of squeezing  of the anti-squeezed quadrature $:\mathrel{S(\omega)_{\frac{\theta+\pi}{2}}}:$ reproduces the profile of the emission spectrum, i.e. a narrow linewidth $\Gamma$ set by the vanishing Liouvillian gap. On the other hand, the spectrum of squeezing of the squeezed quadrature, $:\mathrel{S(\omega)_{\frac{\theta}{2}}}:$, shows a similarly narrow feature that evolves from negative to positive values in a timescale $~\sim \kappa^{-1}$, on top of a broader, negative profile with a linewidth $\sim \kappa$. When integrated in frequencies, this broader negative profile is the main contribution to the corresponding intracavity quadrature fluctuation, which fulfills $\langle :\mathrel{\Delta X_{\phi}^2}: \rangle  =1/(2\pi\kappa)\int_{-\infty}^{\infty} :\mathrel{S(\omega)_\phi }: d\omega$. All these features are consistent with the results observed in the emission from an OPO~\cite{walls_book94a,gardiner_book00a}: the spectrum of the squeezed quadrature has a bandwidth $\sim \kappa$, while the spectrum of anti-squeezed quadrature has a linewidth that vanishes as one approaches the critical point, as so does the emission spectrum. We highlight again the important difference between the squeezed laser and an OPO. The mechanism of stimulated emission, responsible for the vanishing linewidth, plays a negative role in the OPO, since it tends to turn the squeezed vacuum into a mixture of coherent states and thus degrades squeezing. In the squeezed laser, the coherent population is created in the squeezed basis, and the final result is a coherent-squeezed state with the narrowband spectral properties of a laser.

\subsection{Second-order correlation function}
Another common approach to quantify photonic correlations is the use of the second-order standard correlation function, defined in the stationary state as
\begin{equation}
g^{(2)}(\tau) = \lim_{t\rightarrow \infty}\frac{\langle \hat a^\dagger (t) \hat a^\dagger (t+\tau) \hat a (t+\tau) \hat a(t)\rangle }{\langle \hat a^\dagger \hat a (t)\rangle\langle \hat a^\dagger \hat a (t+\tau)\rangle}.
\end{equation}
The introduction of this quantity and related ones by Glauber~\cite{Glauber1963}, in the context of the development of the first lasers, established the notions of quantum-optical coherence and our theoretical understanding of ideal lasers as states that display coherence to all orders, meaning that individual photons within the laser are statistically independent.  Second-order coherence thus translates to $g^{(2)}(\tau) = 1$.

The lasing system that we present here extends the definition of a laser in terms of Glauber's criterion of coherence. While the squeezed laser is a source of extremely monochromatic light with a characteristic coherence time $\tau_c = 1/\Gamma $ that tends to infinity, photons emitted by the squeezed laser exhibit positive correlations between each other, characterized by a value of the zero-delay, stationary second order correlation function $g^{(2)}(0)$ greater than one. 
\begin{figure}[b]
\centering

\includegraphics[width=0.99\columnwidth]{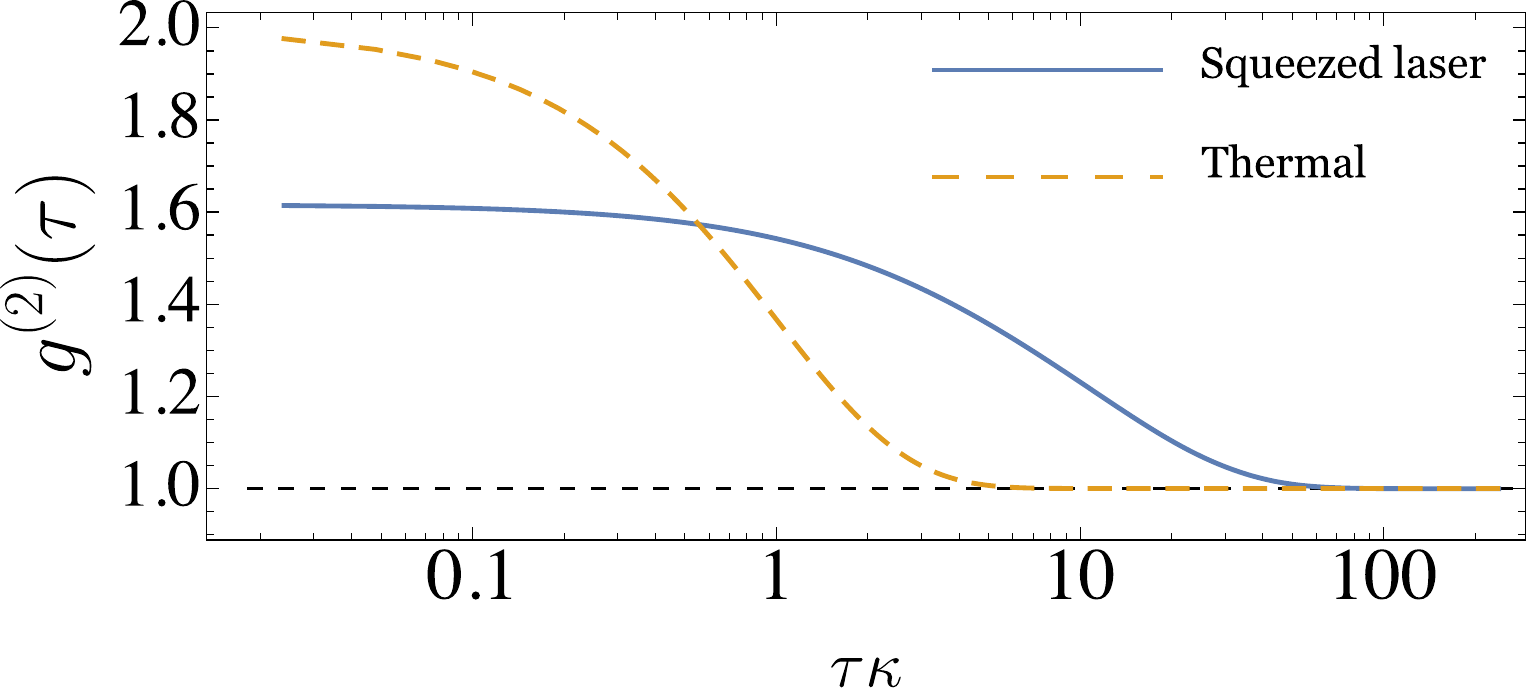}
\caption{$g^{(2)}(\tau)$ of the squeezed laser displaying positive correlations surviving for extremely long correlation times $\propto 1/\Gamma$. A thermal state, showed for comparison, displays a much shorter positive correlation time of the order $1/\kappa$.  Parameters: $C_s = 1.5$, $r=1$, $n_q = 50$.  }
\label{fig:g2}
\end{figure}  

Considering the steady state given by Eq.~\eqref{eq:rho_a}, in the deep-lasing regime ${n_s \gg 1}$,  ${g^{(2)}(0)}$ reads ${g^{(2)}(0) \approx {3-\sech(2r)^2}/2}$. 
which increases with the squeezing parameter $r$ and saturates to a value $3/2$.  For the symmetry-broken states $\hat \rho_\varphi$ considered above, this value can be made even larger.
%The good agreement with exact, numerical results is shown in Fig.~\ref{fig:g2}(a). 
Notably, and contrary to the case of e.g. a thermal state, these positive correlations exhibit an extraordinarily long coherence time $\tau_c$, that can be made orders of magnitude longer than the natural lifetime of the cavity photons, $1/\kappa$, as shown in Fig~.~\ref{fig:g2}(a). These extremely long-lived positive correlations establish the squeezed laser as a novel source of light set apart from standard lasers or thermal sources. 
%In the following, we discuss some of the implications that these properties may have for prospective technological applications. 

\section{Applications in quantum metrology}
Here we briefly discuss the prospects of the squeezed laser for applications in quantum metrology. We first focus on one one of the main applications of squeezed states in metrology: the determination of small phase shifts by some interferometric process. A prominent example is the detection of gravitational waves through a Michelson  interferometer (MI)~\cite{aasi13a,Lvovsky2015,Schnabel2017}. The  application of a squeezed laser in these types of interferometers would require a slightly different measurement approach than the one used with squeezed vacuum, as we now discuss.

\begin{figure}[h]
\begin{center}
\includegraphics[width=0.99\columnwidth]{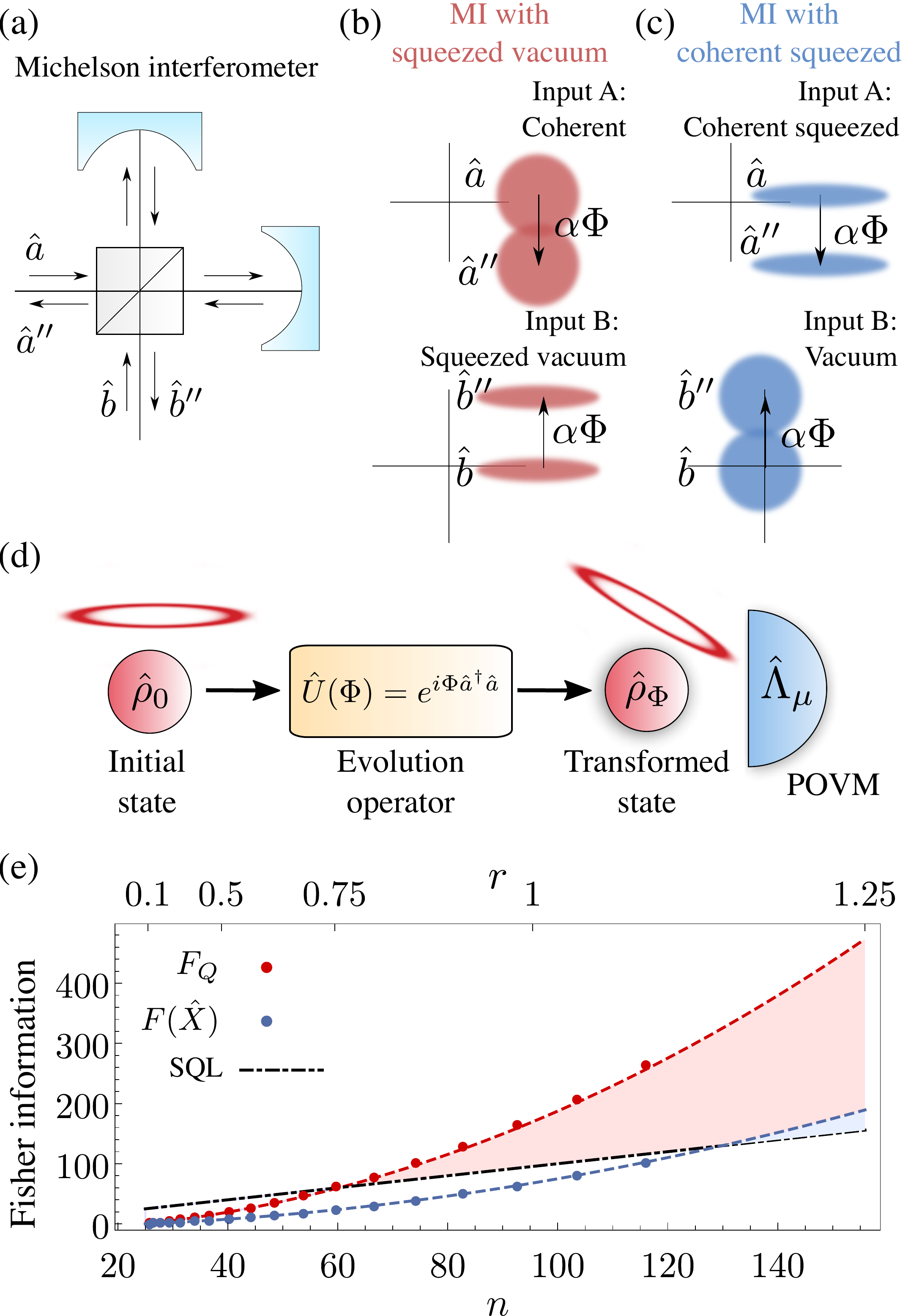}
\end{center}
\caption{Michelson interferometry with a squeezed laser. (a) The two inputs (outputs) of the interferometer are described by the operators $\hat a$, $\hat b$ ($\hat a''$, $\hat b''$). (b) Operation of a MI with squeezed vacuum. (c) Operation of a MI with a squeezed laser. (d) Sketch of the framework of quantum parameter estimation. (e) Quantum Fisher information (red) and classical Fisher information (blue) for $\hat X$ measurement for the estimation of $\Phi$ in the MI with a phase-diffused squeezed laser. Numerical data points are fitted by the quadratic expression in Eq.~\eqref{eq:Fisher-estimation} (dashed lines), thus showing Heisenberg scaling. The dotted-dashed line represents the standard quantum limit.}
\label{fig:michelson}
\end{figure}

The standard application of squeezed states in a MI uses a beam splitter to mix a coherent state, injected through an input port $a$, and a squeezed vacuum, injected through an input port $b$ (see Fig.~\ref{fig:michelson}(a)). Let us consider a beam splitter that implements the transformation ${a'\mathrel{=}(a+b)/\sqrt{2}}$ and ${b' = (a-b)/\sqrt{2}}$. The purpose of the MI to measure a phase shift $\varphi$ on mode $b'$ due to the difference in path length. After recombination back at the beam splitter, the output modes of the interferometer are given by $a''=(a'+b')/2$ and $b''=(a'-e^{i\varphi}b')/2$. In the dark fringe configuration ${e^{i\varphi}\approx 1 + i\varphi}$, we find that $\hat a''\approx(1+i\varphi/2)\hat a - (i\varphi/2) \hat b$ and $\hat b'' \approx -(i\varphi/2)\hat a + (1+i\varphi/2)\hat b$.

In the usual MI scenario, one assumes that $\hat a$ contains a very large coherent component, so that  $\hat a = \alpha + \delta \hat a$, with $\langle \delta\hat a\rangle = 0$ and $\langle \delta\hat a^\dagger \delta\hat a\rangle \ll |\alpha|^2$, and that $\hat b$ is either in vacuum or in a squeezed vacuum. In any case, one can rewrite $\hat b'' \approx \hat b -i\alpha \varphi/2$, provided $\alpha$ is much larger than the fluctuations in $\hat b$, $\alpha \gg \langle \hat b^\dagger \hat b\rangle$. We thus see that the phase $\varphi$ is encoded as a displacement $\varphi \alpha$ of mode $b$, which allows one to benefit from the reduced fluctuations if $b$ is in a squeezed vacuum, see Fig.~\ref{fig:michelson}(b).

However, when using a squeezed laser, both the large coherent component and the squeezed fluctuations are in mode $\hat a$, while $\hat b$ is a standard vacuum. This means that measuring $\hat b''$ will not provide any squeezing-based enhancement. However, we prove here that, by measuring $\hat a''$ instead, one recover the same enhanced sensitivity. Indeed, ignoring the very small term $i\varphi/2\hat b$, we have $\hat a'' \approx (1-i\Phi/2)\hat a \approx e^{-i\Phi/2}\hat a$, since $\Phi \ll 1$. Therefore, $\Phi$ is encoded in a rotation in phase space of mode $\hat a$ of angle $\Phi/2$. In order to keep the connection with the standard MI approach, it is easy to see that this rotation can alternatively be viewed as a small displacement of $i \alpha\Phi/2$ provided $\alpha\gg \langle \delta \hat a^\dagger \delta \hat a\rangle$. The measurement of this small displacement thus benefits from the squeezed quadrature fluctuations in $\delta \hat a$, see Fig.~\ref{fig:michelson}(b).

The case of the MI just discussed is a particular example in which a unitary transformation, ${\hat U_\Phi = \exp\left[i\Phi \hat G/2\right]}$ is applied to some initial state to yield a final state that encodes the phase information,  $\hat \rho_\Phi  = \hat U_\Phi \hat  \rho_0$, see Fig.~\ref{fig:michelson}(d). In the framework of quantum parameter estimation, one considers a positive operator-valued measure (POVM) $\Lambda=\{\hat\Lambda_\mu\}$, whose measurement outcomes follow a probability distribution $p_\Phi(\mu)=\mathrm{Tr}[\rho_\Phi \Lambda_\mu]$. When $\Phi$ is inferred from $p_\Phi(\mu)$, the Cramér-Rao bound establishes that the best attainable sensitivity $\Delta\Phi$ that can be achieved is given by the Fisher information $F(\Lambda)=\mathrm E[(d\log p_\Phi(\mu)/d\Phi)^2]$, so that $\Delta\Phi \geq 1/\sqrt{F(\Lambda)}$~\cite{Dowling2015}. 
Choosing a POVM $\Lambda$ that maximizes $F(\Lambda)$ yields the so-called quantum Fisher information $F_Q$, which only depends on $\hat \rho_\Phi$. If $\hat \rho_\Phi$ is generated by applying some $\hat U_\Phi$ to a pure initial state $\hat \rho_0$, the quantum Fisher information simply reads $F_Q = \Delta \hat G^2$~\cite{Paris2009,Dowling2015}.

As discussed above, the appropriate $\hat G$ that describes the operation of the MI is $\hat G =\hat a^\dagger \hat a$ (i.e. a rotation in phase space), which can be simplified to a displacement $\hat G = \alpha^* \delta\hat a + \alpha \hat \delta a^\dagger$ for very large values of $\alpha$. In any of these cases, a coherent state yields $F_Q = |\alpha|^2 = n$, which sets the standard quantum limit (SQL) of precision $\Delta\Phi\mathrm{SQL} = 1/\sqrt{n}$. The main goal of quantum metrology is to surpass that limit. It is easy to show that a coherent squeezed state $D(\alpha)\hat S(r e^{i\theta})|0\rangle$ can do so, since $F_Q = \Delta n^2 = |\alpha \cosh r - \alpha^* e^{i\theta}\sinh r|^2 + 2\sinh^2 r\cosh^2 r $. For a phase-squeezed state $\theta = \pi +\arg\alpha$, one has $\Delta n^2 = |\alpha|^2e^{2r} + 2\sinh^2r\cosh^2 r$, so that $\Delta n^2 > n = |\alpha|^2+\sinh^2 r$ for many choices of $\alpha$ and $r$, thus breaking the SQL. It is even possible to attain a quadratic scaling of the Fisher information with photon number---the so called Heisenberg limit---, since in the limit of $\sinh^2 r\gg |\alpha|^2$, $F_Q \approx 2\sinh^4 r \approx 2 n^2$. 

It is also interesting to consider whether such a precision beyond the SQL could also be obtained with the phase-diffused stationary state in Eq.~\eqref{eq:rho_a}.  This would correspond to a situation in which the measurement time is much longer than $1/\Gamma$, or in which the measurement involves average over many distinct realizations with no phase coherence between them. In the case of a standard laser, this interferometric process would no longer work, since the phase-diffused state is $U(1)$ symmetric and thus it does not vary under rotations. However, as we show in Fig.~\ref{fig:michelson}(e), even in this phase-diffused situation, the squeezed laser still provides a phase sensitivity that can surpass the SQL and even reach the Heisenberg limit. This is shown through the analysis of both the quantum Fisher information $F_Q$, and the Fisher information $F(\hat X)$ associated with a measurement of $\hat X = (\hat a+\hat a^\dagger)$. We show how both depend on photon population $n$ of the initial state $\rho_0$. $n$ is given by Eq.~\eqref{eq:na_squeezed}, and can be increased by increasing both $n_s$ and $r$. 
Heisenberg scaling is achieved if  $n$ is increased by setting $n_s$ constant and increasing $r$. We have numerically confirmed that,  in that case, the Fisher information is well approximated by
%$^{88}$Sr
\begin{equation}
F \approx \frac{1}{\beta n_s}(n^2 - n_s^2),
\label{eq:Fisher-estimation}
\end{equation} 
with $\beta=2$ for $F_Q$, and $\beta=5$ for $F(\hat X)$.  Further details about this calculation can be found in the Supplemental Material. 

Beyond optical interferometry, the generation of narrowband squeezed light with a squeezed laser could open new doors to the investigation of atomic metrology with squeezed states. Many previous works have discussed the possibility of generating spin squeezing through the coupling between spin ensembles and squeezed states of light~\cite{Hald1999,Kuzmich1997,Hammerer2010,Appel2008,
Honda2008,Tanimura2006,Hetet2007}, and its applications for atomic interferometry~\cite{Agarwal1996,Szigeti2014,Pezze2018} and magnetometry~\cite{Wolfgramm2010,Wolfgramm2013,Horrom2012}. 
Furthermore, quantum properties of light have been proposed as a novel control knob for nonlinear spectroscopy~\cite{Dorfman2016}. In the field of multi-photon microscopy, the use of non-classical light with entangled or strongly-correlated photons such as squeezed states has been proposed theoretically~\cite{Gea-Banacloche1989,Fei1997} and confirmed experimentally~\cite{Georgiades1995,Upton2013,Villabona-Monsalve2017,Li2019} in order to strongly increase two-photon absorption probability and reduced the require intensity of the illuminating field. All the  applications above benefit greatly from a narrowband source of quantum-correlated light able to match narrow atomic linewidths, such as the squeezed laser proposed in this work.

\section{Experimental implementation}
There are two main constraints for the experimental implementation of the squeezed laser. The first one is  imposed by the requirement for a squeezed vacuum with a bandwith much larger than the cavity linewidth, so that it behaves effectively as a squeezed white noise. The second one is the condition $\sqrt{N}g\sinh r \ll \Delta_s$, which is required to apply a rotating-wave approximation and obtain a Jaynes-Cummings type of coupling between the atoms and the squeezed cavity mode. Therefore, if one desires to achieve a specific value of  $r$ (determined by the ratio $\Omega_p/\Delta_c$), one needs to increase $\Delta_c$ and $\Omega_p$ by the same factor until the condition is fulfilled, which might be ultimately limited by the maximum possible power available for the pump and by the free spectral range of the cavity, which should remain larger than the required detuning.

To address these questions, we will take as a reference example recent experiments on superradiant lasing in cold strontium~\cite{Norcia2016,Schaffer2020}. Here, lasing occurs in a long-lived ($\gamma = $~\SI{7.5}{\kHz}) dipole-forbidden $^3 $P$_1\rightarrow ^1 $S$_0$ transition in $^{88}$Sr, corresponding to a frequency $\omega_c/2\pi =$~\SI{435}{\THz} and  with an effective, collective light-matter coupling between atoms and cavity mode $\sqrt N g/2\pi \sim$~\SI{1}{\MHz}~\cite{Norcia2016}. We consider cavities with a linewidth going from $\kappa/2\pi = $~\SI{160}{\kHz} as reported in Ref.~\cite{Norcia2016}, to the more typical values of  $\kappa/2\pi \sim$~\SI{2}{\MHz} common in cavity-QED experiments~\cite{Hamsen2017}. 

The requirement of  squeezed noise with a broad enough bandwith can be met with several, broadband squeezed vacua with bandwidths much larger than \SI{}{\MHz} that have been reported experimentally, with examples ranging from \SI{65}{\MHz} in ring optical parametric oscillators~\cite{Serikawa2016} to \SI{1.2}{\GHz} in monolithic PPKTP cavities~\cite{Ast2013}. 

Regarding the RWA condition, consider that we require a squeezed laser with a value of $r=1$, which in the case of a squeezed vacuum corresponds to a reduction of quadrature variance of $\sim $~\SI{8.5}{\dB}. This value of $r$ is obtained by setting $\alpha \approx 0.46$. 
The RWA  will only hold if $ \Delta_c\sqrt{1-\alpha^2}\approx 0.88 \Delta_c \gg \sqrt N g \sinh(1)$, which is fulfilled for $\Delta_c \gg \sqrt N g$. Assuming  $\Delta_c/2\pi =20\times g/2\pi \sim $~\SI{20}{\MHz}, we would require $\Omega_p \sim$~\SI{10}{\MHz}. In a doubly-resonant situation in which the cavity is resonant with both pumping and signal fields, $\Omega_p$ is related to the power $P$ of the pump by 
$P = A_p (\Omega_p/\kappa)^2 P_0 $~\cite{Patera2010}, where $A_p$ is the waist area of the pump beam, and $P_0$ is given by
\begin{equation}
P_0 = \frac{\epsilon_0 c^3 n_0^2 T_p T_s^2}{32 (\chi l \omega_c)^2},
\end{equation}
with $n_0$, $l$ and $\chi$  the index of refraction, length and nonlinear susceptibility  of the nonlinear crystal, $\omega_c$ the frequency of the cavity mode, and $T_{s(p)}$ the cavity transmission factor at the signal (pump) frequencies. Using typical parameters for a PPKTP crystal, $l = $~\SI{0.5}{\mm}, $\chi =$~\SI{14}{\pico \meter/\V} and $n_0=1.8$~\cite{Lvovsky2015}, and taking $T_p=T_s=T\approx 1.3 \times10^{-4}$ consistent with a free spectral range $\omega_\mathrm{FSR}/2\pi\approx$~\SI{2.6}{\GHz} and $\kappa/2\pi =$~\SI{160}{\kHz}~\cite{Norcia2016}, and a waist radius of the pump beam of \SI{30}{\micro\meter}, the resulting required power would be $P\sim $~\SI{1.5}{\uW}.
Even in a  worse scenario in which the cavity had transmissivity $T\sim 10^{-2}$ and $\kappa/2\pi \approx$~\SI{2}{\MHz} ($\omega_\mathrm{FSR}/2\pi\approx  $~\SI{400}{\MHz}), the required power would only raise to $P\approx$~\SI{4.6}{\mW}. In any of these situations, the required pumping power would remain within reasonable values not above the order of \SI{}{\mW}, and the free spectral range of the cavity would remain smaller than the required detuning $\Delta_c \sim$~\SI{20}{\MHz}.
\section{Conclusions}
 We have introduced the concept of a squeezed laser, and discussed how it could be implemented in an optical setup. Apart from the extremely narrow linewidth and long coherence times typical of any laser, the squeezed laser features strong photon-photon correlations and squeezed quadratures. This represents a new paradigm of lasing in stark contrast to the standard definition of a laser in terms of Glauber's criterion of optical coherence. We have shown that this novel source of light can be applied in Michelson interferometry for metrology beyond the standard quantum limit, and argued that the combination of its squeezed character and narrow linewidth can find important applications in the field of atomic metrology.
 \\[10pt]
 
\acknowledgements
%\paragraph*{Acknowledgements} 
C.S.M. thanks C. Navarrete-Benlloch and D. Martin-Cano for insightful discussions and advice. Both authors acknowledge funding from the European Research Council under the European Union’s Seventh Framework Programme (FP7/2007-2013)/ERC Grant Agreement No. 319286 (Q-MAC).   C. S. M. acknowledges that the project that gave rise to these results received the support of a fellowship from la Caixa Foundation (ID 100010434), from the European Union's Horizon 2020 Research and Innovation Programme under the Marie Sklodowska-Curie Grant Agreement No. 47648, with fellowship code  LCF/BQ/PI20/11760026, and from a Marie Sklodowska-Curie Fellowship QUSON (Project  No. 752180). DJ acknowledges funding from the EPSRC Hub in Quantum Computing and Simulation (EP/T001062/1).

\let\oldaddcontentsline\addcontentsline% Store \addcontentsline
\renewcommand{\addcontentsline}[3]{}% Make \addcontentsline a no-op
\bibliography{sci-url,library,books}

\let\addcontentsline\oldaddcontentsline% Restore \addcontentsline

%\end{document}

\clearpage
\onecolumngrid
\appendix
\renewcommand\appendixname{SM}

\begin{center}
{\bf \large Supplementary Material}
\end{center}

\renewcommand{\theequation}{S\arabic{equation}}

\renewcommand{\thefigure}{S\arabic{figure}} 
\setcounter{figure}{0} 
\setcounter{equation}{0}   

\section{Diagonalization of the Hamiltonian}
%\subsection*{Hamiltonian}
Let us consider the Hamiltonian of a parametrically-driven cavity:
\begin{equation}
H=\Delta_{c}a^{\dagger}a+\frac{\Omega_{p}}{2}\left(e^{i\theta_{p}}a^{2}+\mathrm{h.c.}\right).
\end{equation}
In order to diaonalize this Hamiltonian,we perform a change of basis given by the squeezing unitary
transformation $S(\xi)=\exp\left[\frac{1}{2}(\xi^{*}a^{2}-\xi a^{\dagger2})\right]$
, with $\xi=re^{-i\theta}$, such that
\begin{equation}
S^{\dagger}(re^{-i\theta})aS(re^{-i\theta})=a\cosh r-a^{\dagger}e^{-i\theta}\sinh r.
\end{equation}
We therefore perform the substitution $a  \rightarrow a_{s}\cosh r-a_{s}^{\dagger}e^{-i\theta}\sinh r$ in the Hamiltonian,  where $a_{s}$ is describes the annihilation operator
in the new ``squeezed'' basis.  This gives
have
\begin{multline}
H=\left\{ a_{s}^{2}\left[\frac{\Omega_{p}}{2}\left(e^{i\theta_{p}}c^{2}+s^{2}e^{-i(\theta_{p}-2\theta)}\right)-\Delta_{c}e^{i\theta}sc\right]+\mathrm{h.c.}\right\} +a_{s}^{\dagger}a_{s}\left[-\Omega_{p}cs\left(e^{i(\theta_{p}-\theta)}+e^{-i(\theta_{p}-\theta)}\right)+(c^{2}+s^{2})\Delta_{c}\right]\\
-\frac{\Omega_{p}}{2}cs\left(e^{i(\theta_{p}-\theta)}+e^{-i(\theta_{p}-\theta)}\right)+\Delta_{c}s^{2},
\end{multline}
where we used $c=\cosh r$ and $s = \sinh r$.  In order for the Hamiltonian to be diagonal, $r$ and $\theta$ must be such that 
\begin{equation}
\frac{\Omega_{p}}{2}\left(e^{i\theta_{p}}c^{2}+s^{2}e^{-i(\theta_{p}-2\theta)}\right)-\Delta_{c}e^{i\theta}sc=0.
\end{equation}
The imaginary part of this equation gives
\begin{equation}
\sin(\theta_{p}-\theta)\frac{\Omega_{p}}{2}\left(1-\tanh^{2}r\right)=0,
\end{equation}
which sets the condition $\theta=n\pi+\theta_{p}$. This yields $\cos(\theta_{p}-\theta)=\pm1$ depending on the parity
of $n$. Therefore, for the real part:
\begin{equation}
\pm\frac{\Omega_{p}}{2}\left(1+\tanh^{2}r\right)-\Delta_{c}\tanh r=0
\end{equation}
Choosing $\theta=\theta_{p}$, we select the positive value, and defining
$\alpha=\Omega_{p}/\Delta_{c}$, we obtain:
\begin{equation}
\tanh r=\frac{1-\sqrt{1-\alpha^{2}}}{\alpha}.
\end{equation}
Using $\tanh^{-1}x=\frac{1}{2}\ln\frac{1+x}{1-x}$, we find:
\begin{equation}
r=\frac{1}{2}\ln\frac{\alpha+1-\sqrt{1-\alpha^{2}}}{\alpha-1+\sqrt{1-\alpha^{2}}}=\frac{1}{4}\ln\frac{1+\alpha}{1-\alpha}.
\end{equation}

\section{Derivation of conditions for the squeezed drive}
In this section, we derive the conditions required for the squeezed drive in order to cancel squeezed noise in the squeezed basis. We start by considering the master equation of a system driven by a broadband squeezed white
noise with squeezing parameter $r_{e}e^{i\theta_{e}}$. The coupling to the squeezed reservoir has an associated decay rate $\kappa$. Other channels of decay (by coupling to different baths in vacuum, e.g. decay through a second mirror, or intra-cavity losses) have an associated decay rate $\eta \kappa$, with $\eta$ a dimensionless factor.  The resulting master equation reads
\begin{equation}
\partial_{t}\rho=-i[H,\rho]+\frac{\kappa}{2}(N+1+\eta)L_{a}[\rho]+\frac{\kappa}{2}NL_{a^{\dagger}}[\rho]-\frac{\kappa M}{2}L'_{a}[\rho]-\frac{\kappa M^{*}}{2}L'_{a^{\dagger}}[\rho],
\end{equation}
with
\begin{align}
L_{O}[\rho] & \equiv2O^{\dagger}\rho O-O^{\dagger}O\rho-\rho O^{\dagger}O,\\
L'_{O}[\rho] & \equiv2O\rho O-OO\rho-\rho OO,
\end{align}
and
\begin{align}
M & =\cosh(r_{e})\sinh(r_{e})e^{-i\theta_{e}},
\label{eq:M}
\\
N & =\sinh^{2}(r_{e}).
\label{eq:N}
\end{align}

\begin{figure}[t]
\begin{center}
\includegraphics[width=0.5\columnwidth]{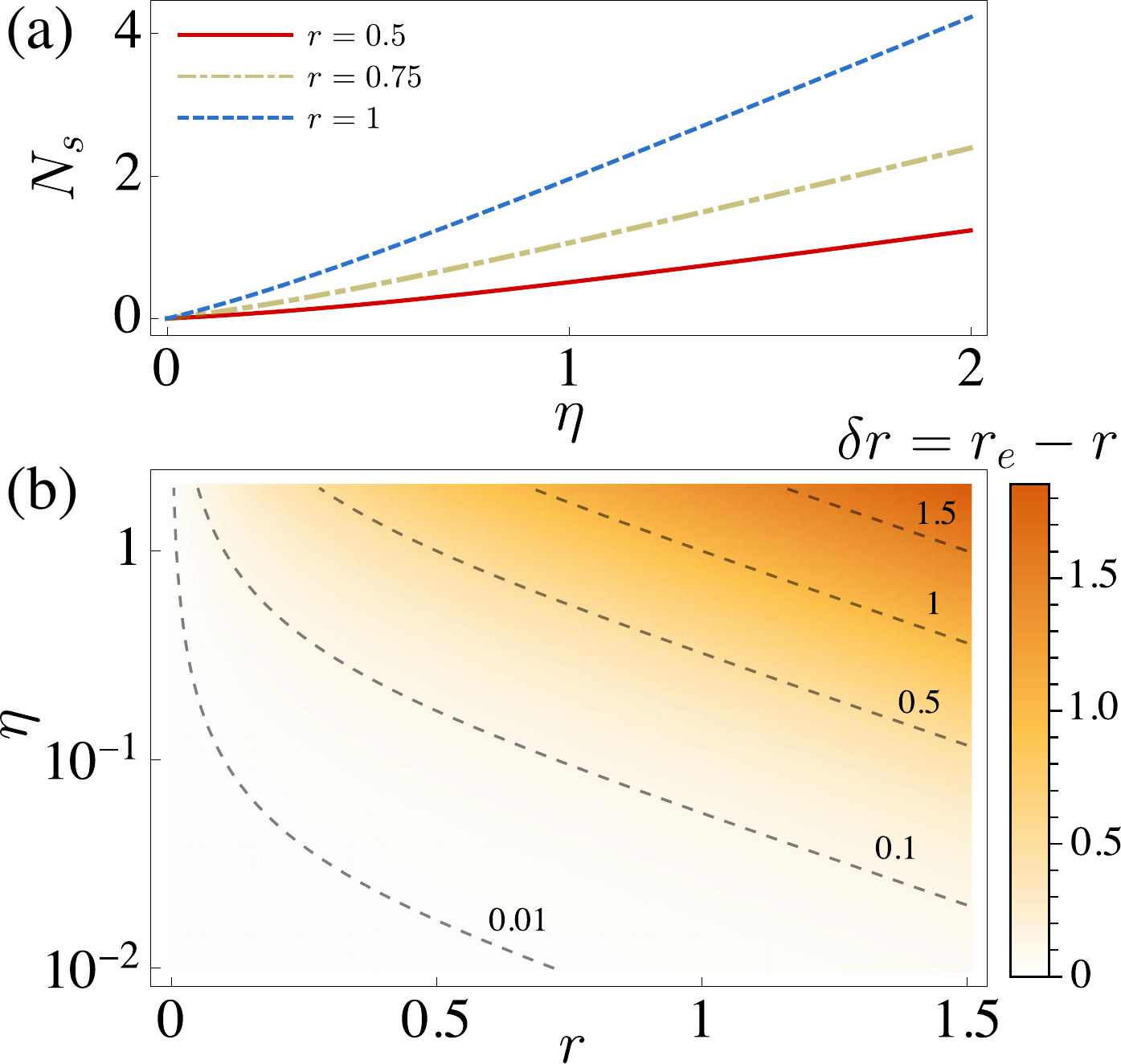}
\end{center}
\caption{(a) Effective thermal photon number versus $\eta$ for different values of $r$. (b) Difference between the squeezing factor $r$ generated inside of the cavity and the required $r_e$ for the external field to cancel squeezed noise.  
}
\label{fig:intracavity-loss}
\end{figure}

Applying the squeezing transformation that diagonalizes the Hamiltonian as discussed above,  the master equation is transformed in the following way
\begin{equation}
\partial_{t}\rho=\frac{\kappa}{2}\left(1+\eta+N_{s}\right)L_{a_{s}}[\rho]+\frac{\kappa}{2}N_{s}L_{a_{s}^{\dagger}}[\rho]-\frac{\kappa}{2}M_{s}L'_{a_{s}}[\rho]-\frac{\kappa}{2}M_{s}^{*}L'_{a_{s}^{\dagger}}[\rho]
\label{eq:master-squeezed}
\end{equation}
where, again, we used $c\equiv \cosh r$ and $s \equiv \sinh r$, and
\begin{align}
N_{s} & =s^{2}(1+\eta)+N(s^{2}+c^{2})+cs(Me^{-i\theta}+M^{*}e^{i\theta})\\
M_{s} & =cse^{i\theta}(2N+1+\eta)+Mc^{2}+M^{*}e^{2i\theta}s^{2}.
\end{align}
From Eq.~\eqref{eq:master-squeezed} one can  understand the problem inherent to the dynamics in the squeezed basis: normal decay turns into thermal and squeezed noise. Our inclusion of a squeezed drive allows to correct for this problem, i.e., it allows us to select $N$ and $M$ so that $N_{s}$ and $M_{s}$ are equal to zero. This way, one 
recovers a standard decay term in the squeezed basis. 
Substituting back the values of $N$ and $M$ from Eq.~\eqref{eq:M} and \eqref{eq:N}, we obtain:
\begin{align}
N_{s} & =s^{2}(1+\eta)+(s^{2}+c^{2})\sinh^{2}r_{e}+2cs\cosh(r_{e})\sinh(r_{e})\cos(\theta+\theta_{e})\\
M_{s} & =cse^{i\theta}(2\sinh^{2}r_{e}+1+\eta)+\cosh(r_{e})\sinh(r_{e})c^{2}e^{-i\theta_{e}}+s^{2}\cosh(r_{e})\sinh(r_{e})e^{i(\text{\ensuremath{\theta_{e}}+}2\theta)}.
\end{align}
Setting $M_{s}=0$, and taking the imaginary part, we get
\begin{equation}
\sin(\theta+\theta_{e})\cosh(r_{e})\sinh(r_{e})=0,
\end{equation}
which sets the condition for the angles:
\begin{equation}
\theta_{e}+\theta=n\pi\quad n=0,1,2,\ldots
\end{equation}
Considering an odd $n$, so that $\cos(\theta_{e}+\theta_{p})=-1$,
the real part reads:
\begin{equation}
\mathrm{Re}M_{s}=\cos\theta\left[cs(2\sinh^{2}r_{e}+1+\eta)-\cosh(r_{e})\sinh(r_{e})c^{2}-s^{2}\cosh(r_{e})\sinh(r_{e})\right].
\end{equation}
Equating to zero, we find
\begin{equation}
\sinh[2(r_{p}-r_{e})]+\sinh(2r_{p})\eta=0,
\end{equation}
with solution for $r_{e},$ 
\begin{equation}
r_{e}=r_{p}+\frac{1}{2}\mathrm{arcsinh}\left[\eta\sinh(2r_{p})\right].
\end{equation}
In the case $\eta\ll\sinh(2r_{p})$,  we can write a simpler expression for $r_e$, 
\begin{equation}
r_{e}\approx r_{p}+\frac{1}{2}\eta\sinh(2r_{p}).
\end{equation}
This means that, formally, one can always choose a squeezing factor for the external field, $r_e$, that will cancel all the squeezed-like noise that appear when one moves to the the squeezed basis. The remaining noise will take the form of standard thermal noise, describing the coupling to an effective thermal bath of mean photon
number $N_{s}$:
\begin{equation}
N_{s}=\frac{1}{2}\left[ 2\eta\sinh^{2}r_{p}+\sqrt{1+\eta^{2}\sinh^{2}(2r_{p})}-1\right] \\
\approx\eta\left[\sinh^{2}r_{p}+\frac{1}{4}\eta\sinh^{2}(2r_{p})\right]
\end{equation}
$N_s$ can only be exactly zero for $\eta =0$. However, for reasonable values $\eta \leq 1$, the resulting values of $N_{s}$ are very low. For instance, values of $r=1$, and $\eta=1$, yield a photon thermal number $N_{s}\approx1.4$. Figure~\ref{fig:intracavity-loss}(a) shows the values obtained for $N_s$ versus $\eta$ for several values of $r$, showing that $N_s$ always remains of the order $\sim 1$ for the choices of parameters that we consider in this work. Figure~\ref{fig:intracavity-loss}(b) also shows the difference $\delta r = r_e-r$ (i.e., how much ``extra'' squeezing needs to be provided by the external squeezed field in order to neutralize squeezed noise), showing that, if the extra cavity losses  can be kept small such that e.g. $\eta \sim 0.1$, $r_e-r$ can be smaller than $0.1$ for reasonable values of $r$.

\section*{Calculations of quantum and classical Fisher information} 
In the paradigm of quantum parameter estimation, the aim is to measure a   phase $\Phi$ which has been encoded in a state $\rho_\Phi$. This encoding takes place through a unitary transformation $\hat U_\Phi$, dependent on $\Phi$, applied onto an initially prepared state $\rho_0$. In our case, we consider a transformation given by simple phase rotation, $\hat U_\Phi = \exp[i\Phi a^\dagger a/2]$. 
We then perform a generalized quantum measurement given by a positive operator-valued measure (POVM). A POVM consist of a series of operators $\Lambda_\mu$, with $\mu\in \{1,2,\ldots M\}$ denoting the outcome of the measurement, whose probability of occurring is $p_\Phi(\mu) = \mathrm{Tr}[\rho_\Phi \Lambda_\mu]$. To ensure that $p\varphi(\mu)$ is a valid probability distribution, the elements of the POVM must add up to the identity, $\sum_\mu \Lambda_\mu = \mathbb{1}$. Our goal is to estimate $\Phi$ from the outcome of the measurements.  

For $\nu$ measurement outcomes, the classical Cramer-Crao bound establishes that the minimal mean-square error in our estimation of $\Phi$ is given by $\Delta^2\Phi \geq 1/\nu F(p_\Phi)$, where $F(p_\Phi)=E\left[ -\frac{\mathrm d ^2}{\mathrm d \Phi^2}\log p_\Phi \right]$ is the classical Fisher information of the probability distribution~\cite{Dowling2015,Paris2009,Demkowicz-Dobrzanski2015}. The quantum Fisher information $F_Q$ is an upper bound of the Fisher information that is independent of the POVM and that relies only on the state encoding $\rho_\Phi$ . $F_Q$ is thus defined as the optimum classical Fisher information over all possible POVMs, $F_Q(\rho_\Phi) = \max_{\Lambda_\mu}F(\rho_\Phi,\Lambda_\mu)$. The quantum Fisher information can be obtained as the variance of the symmetric logarithmic derivative $L_\Phi$, an operator defined by the equation $\partial_\Phi \rho = \{L_\Phi,\rho_\Phi \}/2$, so that $F_Q = \mathrm{Tr}[L_\Phi^2 \rho]$~\cite{Paris2009}. 
In the case in which the estimated parameter is encoded in the state via a unitary transformation $U_\Phi = \exp[-i H\Phi]$, the quantum Fisher information can be computed as
\begin{equation}
F_Q = \sum_{i,j}\frac{2|\langle e_i | H |e_j\rangle| ^2 (\lambda_i-\lambda_j)^2}{\lambda_i+\lambda_j}
\label{eq:fisher_Q}
\end{equation}
where $|e_i\rangle$ and $\lambda_i$ are the eigenvalues and eigenvectors of the initial state $\rho_0$~\cite{Demkowicz-Dobrzanski2015}. In this case, the quantum Fisher information is independent of $\Phi$. Our calculations of the quantum Fisher information shown in the main text has been obtained from Eq.~\eqref{eq:fisher_Q}, using a numerical diagonalization of $\rho_0$. For the calculations of the classical Fisher information for a measurement of the $X$ quadrature,  we have computed the probability distribution  for the state $\rho_\Phi$ by integrating the corresponding Wigner function $p(x)=\int dp\, W_\Phi(x,p)$, which we computed numerically. In this case, $\rho_0$ is taken to be the steady state of the squeezed laser with a squeezing angle $\theta$ that maximizes the Fisher information. As $r$ increases, the optimum $\theta$ that gives the largest Fisher information when measuring $X$ tends to $\pi$ (not shown).

\end{document}